\documentstyle[twoside,fleqn,espcrc2,epsf]{article}

\def\lsim{\raise0.3ex\hbox{$<$\kern-0.75em\raise-1.1ex\hbox{$\sim$}}}
\def\gsim{\raise0.3ex\hbox{$>$\kern-0.75em\raise-1.1ex\hbox{$\sim$}}}

\newcommand{\p}{\partial}
\newcommand{\pslash}{p\kern-1ex /}
\newcommand{\Dslash}{{\cal D}\kern-1.5ex /}

\title{\vspace*{-2.cm}
\begin{flushright}
{\normalsize UTHEP-451}\\
{\normalsize UTCCP-P-116}\\
\end{flushright}
Maximum entropy analysis of hadron spectral functions and
excited states in quenched lattice QCD\thanks{Talk presented by T.\ Yamazaki}}

\author{CP-PACS Collaboration :
  S.~Aoki\rlap,\address{Institute of Physics,
    University of Tsukuba, Tsukuba, Ibaraki 305-8571, Japan}
  R.~Burkhalter\rlap,$^{\rm a,}$\address{Center for Computational Physics,
    University of Tsukuba, Tsukuba, Ibaraki 305-8577, Japan}
  M.~Fukugita\rlap,\address{Institute for Cosmic Ray Research,
    University of Tokyo, Kashiwa, Chiba 277-8582, Japan}
  S.~Hashimoto\rlap,\address{High Energy Accelerator Research Organization
    (KEK), Tsukuba, Ibaraki 305-0801, Japan}
  N.~Ishizuka\rlap,$^{\rm a,b}$
  Y.~Iwasaki\rlap,$^{\rm a,b}$
  K.~Kanaya\rlap,$^{\rm a}$
  T.~Kaneko\rlap,$^{\rm d}$
  Y.~Kuramashi\rlap,$^{\rm d}$
  M.~Okawa\rlap,$^{\rm d}$
  Y.~Taniguchi\rlap,$^{\rm a}$
  A.~Ukawa\rlap,$^{\rm a,b}$
  T.~Yamazaki\rlap,$^{\rm a}$ and
  T.~Yoshi\'e$^{\rm a,b}$
}

\begin{document}
\pagestyle{empty}

\begin{abstract}
Employing the maximum entropy method
we extract the spectral functions from meson correlators
at four lattice spacings in quenched QCD with the Wilson quark 
action. 
We confirm that the masses and decay constants,
obtained from the position and the area of peaks,
agree well with the results from the conventional exponential fit.
For the first excited state,
we obtain
$m_{\pi_1} = 660(590)$ MeV, $m_{\rho_1} = 1540(570)$ MeV,
and $f_{\rho_1} = 0.085(36)$ in the continuum limit.
\end{abstract}

\maketitle

\section{Introduction}
\label{sec:intro}
The spectral function $f(\omega)$ of hadrons includes
information such as masses and decay constants for various bound 
states and the continuum spectrum for multi-particle states.
Recently the maximum entropy method~\cite{jarr} (MEM)
has been employed
in order to extract the spectral function from
lattice QCD data~\cite{for,nak}.

In this article we apply the MEM to the quenched lattice QCD data 
for pseudoscalar(PS) and vector(V) mesons
previously calculated at $\beta =$ 5.90, 6.10, 6.25, and 6.47~\cite{CPdata}.
We present the corresponding spectral functions for the case of a point source,
and check the reliability of the MEM by comparing
the masses and decay constants from the spectral function 
with those from the exponential fit.
We then extract 
the masses and decay constants for the first excited state in the 
continuum limit.
For details we refer to ref.~\cite{paper}.

\section{Maximum entropy method}
In the MEM \pagebreak the spectral function is determined 
by maximizing 
the quantity P[$F|DH$],
which is the conditional probability of
the spectral function $F$ for a given data $D$ and all prior knowledge $H$
such as $f(\omega) \ge 0$.
By Bayes's theorem 
this probability is replaced with 
${\mathrm P}[D|FH]{\mathrm P}[F|H],$
where
P[$D|FH$] is proportional
to exp$(-\chi^2/2)$ with $\chi^2$ the standard chi-squared, and 
P[$F|H$] to exp$(\alpha S(f))$
with $\alpha$ a real positive parameter.
Here 
the entropy $S(f)$~\cite{jarr} is defined by $f(\omega)$ and 
a positive function called model.
Combining these two factors,
we can determine the most probable spectral function
as the solution of the equation $\p Q_{\alpha}/\p f = 0$,
where 
$Q_{\alpha} = \alpha S(f) -\chi^2/2$.
Finally the result is averaged over $\alpha$ with 
a weight factor P[$\alpha|DH$].

\begin{figure}[t!]
\centerline{\epsfxsize=6.cm \epsfbox{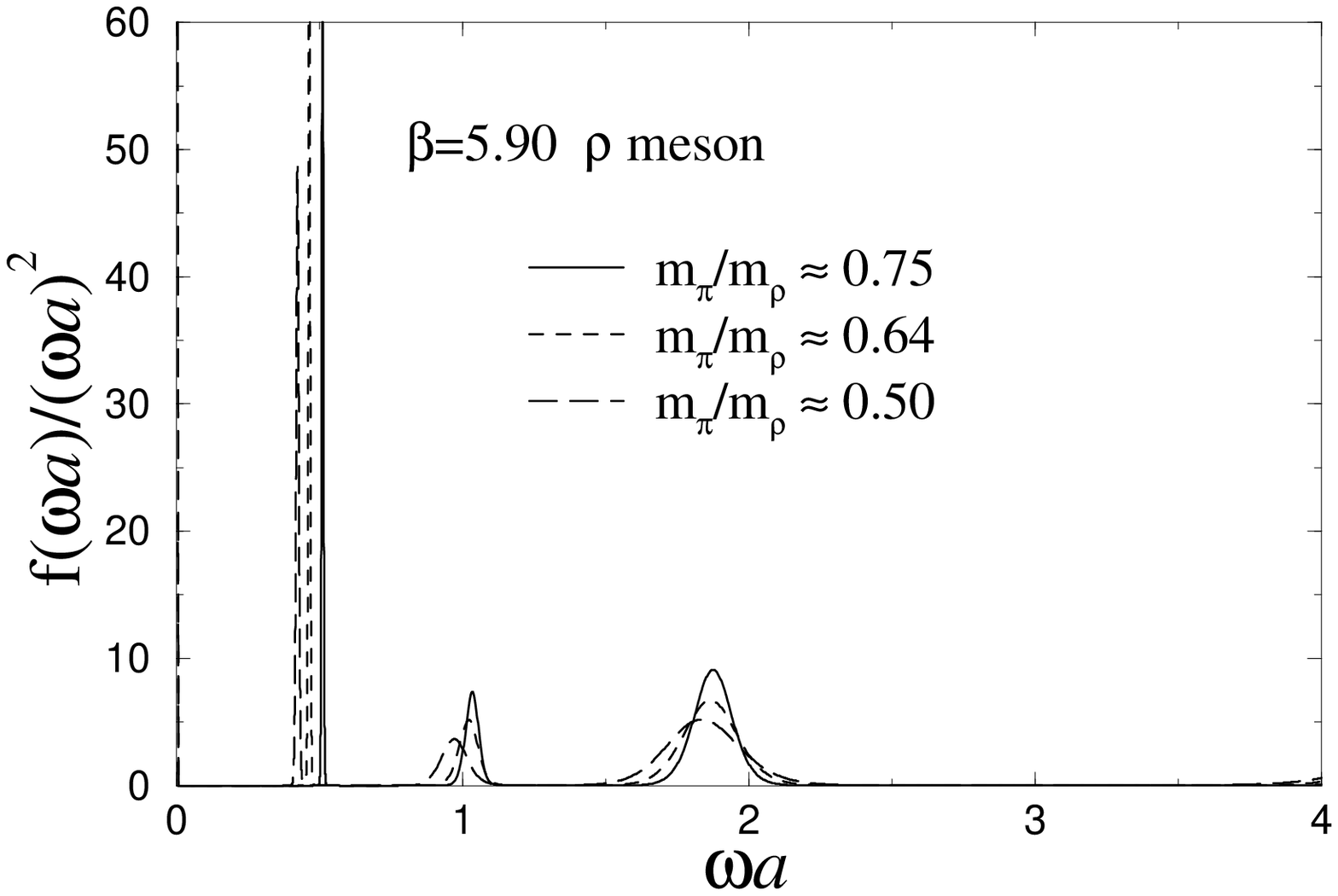}}
\vspace{-0.3cm}
\centerline{\epsfxsize=6.cm \epsfbox{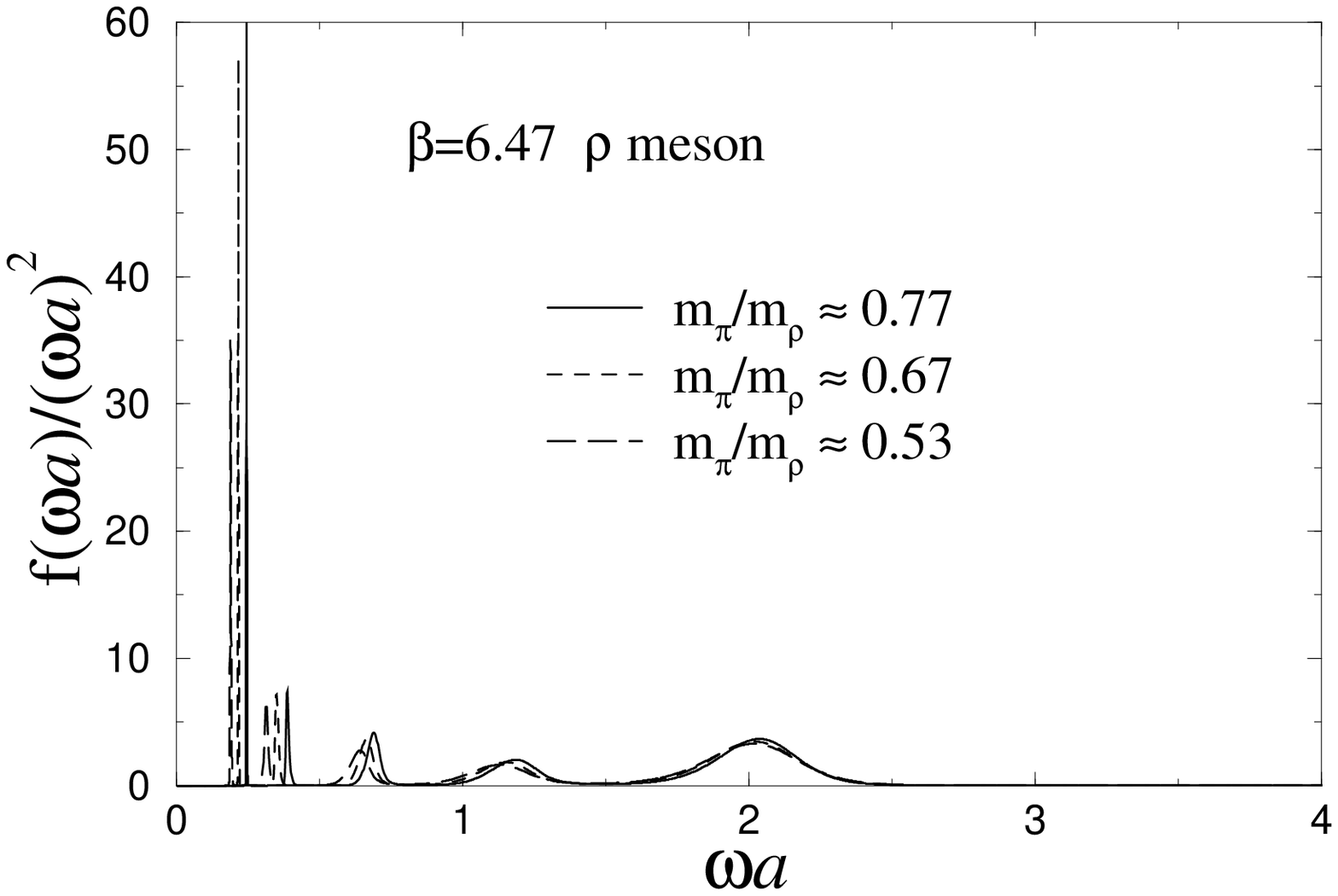}}
\vspace{-1.2cm}
\caption{Spectral functions at $\beta=$ 5.90 and 6.47 for $\rho$ meson.
\label{fig:spec}}
\vspace{-5mm}
\end{figure}

\section{Results}

In Fig.~\ref{fig:spec} a typical result for the spectral functions 
of $\rho$ meson is presented 
at $\beta =$ 5.90 and 6.47 for three different values of $m_{\pi}/m_{\rho}$.
The parameters of the MEM analysis are compiled in ref.~\cite{paper}.
At each $\beta$ one observes, as expected, that
the peaks move to smaller $\omega a$ as quark mass decreases.
As $\beta$ increases, the peaks of the ground and the first excited state 
also move to smaller $\omega a$ as expected. 
In addition the number of peaks increases,
since higher excited states appear below the cutoff $\pi/a$
at larger $\beta$.
While the peak of the ground state is very narrow,
the peaks for the higher states have larger widths,
the reason for which is not understood at present.

\subsection{Mass}
\begin{figure}[t!]
\centerline{\epsfxsize=6.4cm \epsfbox{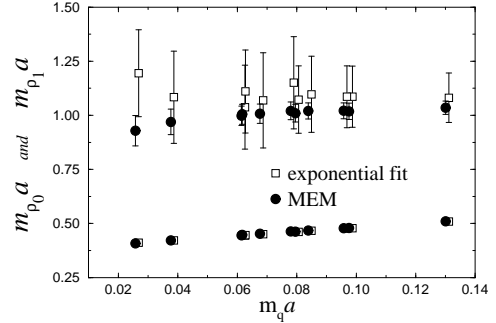}}
\vspace{-1.2cm}
\caption{Comparing the results with these from the exponential fit.
\label{fig:mass}}
\vspace{-7mm}
\end{figure}

The mass of a state is determined from the position of the peak in 
the spectral function.
To check the reliability of this analysis,
the masses of the ground and the first excited state
obtained in this way
are compared with those from the double exponential fit of propagators for
the point and smeared sources.
At $\beta = 5.90$ for $\rho$ meson, 
the two results for the ground state agree,
and those for the first excited state are consistent within errors,
as seen in Fig.~\ref{fig:mass}.

\begin{figure}[t]
\centerline{\epsfxsize=6.2cm \epsfbox{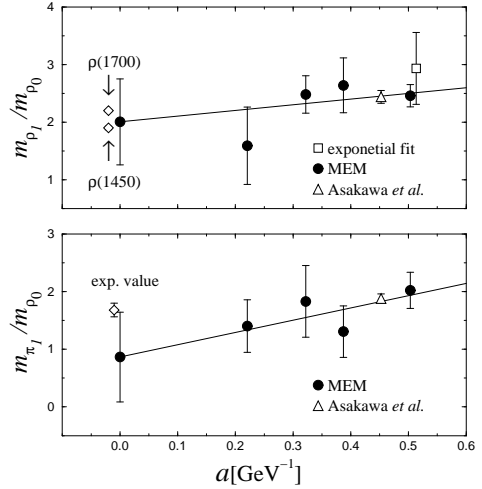}}
\vspace{-1.2cm}
\caption{Continuum extrapolation for the first excited state masses for
both mesons.
\label{fig:masscon}}
\vspace{-5mm}
\end{figure}

The first excited state masses for pseudoscalar and vector mesons 
in the chiral limit at each $\beta$ are
extrapolated to the continuum limit in Fig.~\ref{fig:masscon}.
The error for the vector meson at $\beta =5.90$
is smaller than that obtained by the exponential fit(open square), and
our result is consistent with that of ref.~\cite{nak}(open triangle).
In the continuum limit we obtain
$m_{\pi_1} = 660(590)$ MeV and $m_{\rho_1} = 1540(570)$ MeV,
which are consistent with the experimental values,
though the errors are large.


\subsection{Decay constant}

The decay constant is 
related to the area of the peak in the spectral function~\cite{nak,paper}, 
i.e.,
$f_{\pi}^2\propto 2m_q\!\! \int_{{peak}}
\!\! d\omega\,f_{{\mathrm PS}}(\omega)/m_{\pi}^3$ for pseudoscalar meson and
$f_{\rho}^2\propto \int_{{peak}}
\!\! d\omega\,f_{{\mathrm V}}(\omega)/m_{\rho}^3$ for vector meson.

\begin{figure}[t]
\centerline{\epsfxsize=6.2cm \epsfbox{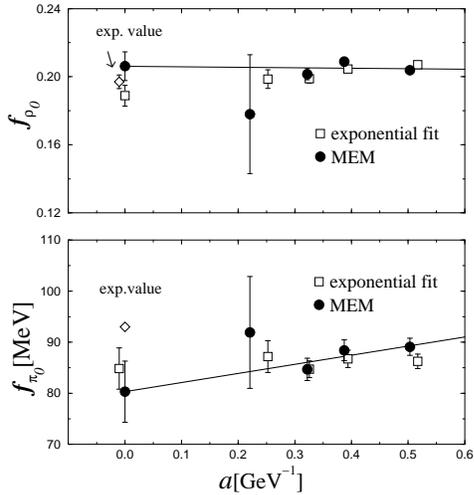}}
\vspace{-1.2cm}
\caption{Decay constant for the ground state for both mesons.
\label{fig:dec0}}
\vspace{-10mm}
\end{figure}

As shown in Fig.~\ref{fig:dec0},
the ground state decay constants for both mesons 
in the chiral limit are
consistent with the results obtained by the exponential fit
at each $\beta$ and in the continuum limit.

\begin{figure}[t]
\centerline{\epsfxsize=6.3cm \epsfbox{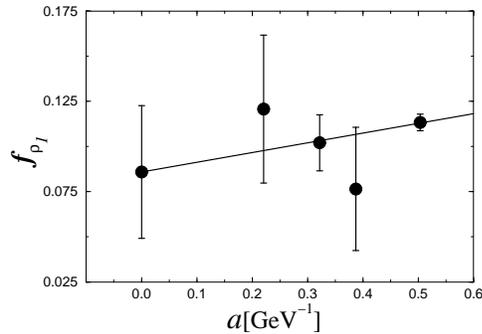}}
\vspace{-1.2cm}
\caption{Decay constant for the first excited state of $\rho$ meson.
\label{fig:decr1}}
\vspace{-5mm}
\end{figure}

The decay constants are determined also for the first excited state.
For the excited $\pi$ meson the decay constant vanishes in the chiral limit,
since $f_{\pi_1}\propto m_{q}/m_{\pi_1}^2$ by definition. 
On the other hand, in Fig.~\ref{fig:decr1}, 
the decay constant for $\rho$ meson is finite in this 
limit, and we obtain $f_{\rho_1} = 0.085(36)$ in the continuum limit.


\subsection{Unphysical state}

\begin{figure}[t]
\centerline{\epsfxsize=6.3cm \epsfbox{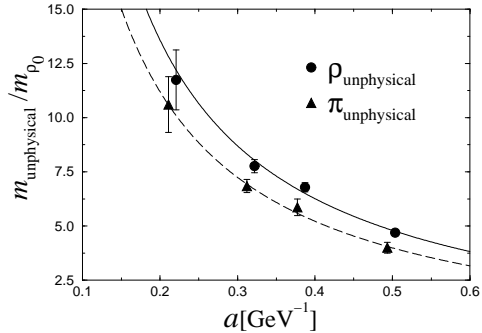}}
\vspace{-12mm}
\caption{Divergent continuum 
extrapolation for the unphysical state mass for both mesons.
\label{fig:unmass}}
\vspace{-5mm}
\end{figure}

We observe in Fig.~\ref{fig:spec} that there is a peak at $\omega a \approx 2$
almost independent of $\beta$.
A similar peak is also present in the pseudoscalar channel.
Apparently the masses of these states in physical units diverge
in the continuum limit, as seen in Fig.~\ref{fig:unmass}.
Here we interpret these states as bound states of
two fermion doublers~\cite{paper}:
In free theory the mass of a two doublers system, each with a 
spatial momentum $\pi a$, is $2\times \log (3)\approx 2.2$, 
which is interestingly close to $\omega a \approx 2$.
The difference may be ascribed to its binding energy.
We note that the bound state of a physical quark and a doubler 
can not appear at zero spatial momentum.

\section{Conclusion}

We have applied the maximum entropy method to our precision quenched lattice
QCD data~\cite{CPdata} 
to extract the spectral functions for pseudoscalar and vector mesons.
The masses and the decay constants extracted from the spectral functions
for the ground state 
are consistent with the ones determined by the exponential fit,
while those for the first excited state 
are new.
In future it will be interesting to apply the MEM analysis to full QCD,
QCD at finite temperature and a study of decay and scattering.

\vspace{2mm}
This work is supported in part by Grants-in-Aid of the Ministry of Education 
(Nos.~10640246, 
10640248, 
11640250, 
11640294, 
12014202, 
12304011, 
12640253, 
12740133, 
and
13640260  
).

%

%
%


\begin{thebibliography}{l}

\bibitem{jarr}M.~Jarrell and J.E.~Gubernatis,
Phys.\ Rep.\ 
{\bf 269} (1996) 133.

\bibitem{for}Ph.~de Forcrand {\it et al.},
Nucl.\ Phys.\ {\bf B} (Proc.\ Suppl.) {\bf 63} (1998) 460.

\bibitem{nak}M.~Asakawa, T.~Hatsuda, and Y.~Nakahara,
Prog.\ Part.\ Nucl.\ Phys.\ {\bf 46} (2001) 459.

\bibitem{CPdata}CP-PACS Collaboration: S.~Aoki {\it et al.},
Phys.\ Rev.\ Lett.\ 
{\bf 84} (2000) 238.

\bibitem{paper}CP-PACS Collaboration: T.~Yamazaki {\it et al.},
hep-lat/0105030.

\end{thebibliography}
\end{document}